\documentclass[12pt]{iopart}
\usepackage[T1]{fontenc}
\usepackage{times}
\usepackage{graphicx}
\usepackage{lineno}
\usepackage{amssymb}
\usepackage{color}
\begin{document}


\title{Acousto-electrical speckle pattern in Lorentz force electrical impedance tomography}
\author{Pol Grasland-Mongrain$^1$, Fran\c cois Destrempes$^1$, Jean-Martial Mari$^2$, R\'emi Souchon$^3$, Stefan Catheline$^3$, Jean-Yves Chapelon$^3$, Cyril Lafon$^3$, Guy Cloutier$^{1,4}$}
\address{(1) Laboratoire de Biorh\'eologie et d'Ultrasonographie M\'edicale, Centre Hospitalier de l'Universit\'e de Montr\'eal (CRCHUM), Tour Viger, 900 rue Saint-Denis, Montr\'eal (QC) H2X 0A9, Canada ;}
\address{(2) GePaSud, University of French Polynesia, Polyn\'esie Fran\c{c}aise ;}
\address{(3) Inserm, U1032, LabTau, Lyon, F-69003, France ; Universit\'e de Lyon, Lyon, F-69003, France ;}
\address{(4) D\'epartement de radiologie, radio-oncologie et m\'edecine nucl\'eaire, et Institut de g\'enie biom\'edical, Universit\'e de Montr\'eal, Montr\'eal (QC) H3T 1J4, Canada}
\ead{pol.grasland-mongrain@ens-cachan.org, guy.cloutier@umontreal.ca}
\noindent{\it Keywords}: Lorentz force, speckle, electrical impedance tomography, magneto-acousto-electrical tomography, electrical conductivity

\submitto{\PMB}

\begin{abstract}
Ultrasound speckle is a granular texture pattern appearing in ultrasound  imaging. It can be used to distinguish tissues and identify pathologies. Lorentz force electrical impedance tomography is an ultrasound-based medical imaging technique of the tissue electrical conductivity. It is based on the application of an ultrasound wave in a medium placed in a magnetic field and on the measurement of the induced electric current due to Lorentz force. 
Similarly to ultrasound imaging, we hypothesized that a speckle could be observed with Lorentz force electrical impedance tomography imaging. In this study, we first assessed the theoretical similarity between the measured signals in  Lorentz force electrical impedance tomography and in ultrasound imaging modalities. We then compared experimentally the signal measured in both methods using an acoustic and electrical impedance interface. Finally, a bovine muscle sample was imaged using the two methods. Similar speckle patterns were observed. This indicates the existence of an ``acousto-electrical speckle'' in the Lorentz force electrical impedance tomography 
with spatial characteristics driven by the acoustic parameters but due to electrical impedance inhomogeneities instead of acoustic ones as is the case of ultrasound imaging.
\end{abstract}



\pagebreak


\section{Introduction}
Ultrasound (US) B-mode is a widespread medical imaging technique of the tissue acoustical impedance. With this technique, an ultrasound wave is transmitted by an acoustic transducer and is backscattered by the change in acoustic impedance at tissue interfaces. Measuring the analytic signal obtained from ultrasound waves received on the transducer surface leads to the reconstruction of these acoustic impedance interfaces. Moreover, when ultrasound waves are scattered by acoustic impedance inhomogeneities, their contributions add up constructively or destructively on the transducer surface. The total intensity thus varies randomly due to these interferences \cite{cobbold2007foundations}. The phenomenon appears on reconstructed images as a granular pattern with variable size and intensity, which looks spatially random: it is called ``acoustic speckle'' \cite{abbott1979acoustic}.

The speckle can be treated as noise and consequently be reduced, or on the contrary be considered as a feature \cite{noble2006ultrasound}. Under the first point of view, quantitative ultrasound methods were proposed based on the backscattering coefficient \cite{lizzi1983theoretical,lizzi1987relationship,insana1990describing}, where the speckle is removed to obtain a measure that depends only on tissue acoustic properties; see \cite{goshal2013state} for further references. Under the second point of view, quantitative ultrasound methods were developed by exploiting the statistics of the echo envelope \cite{burckhardt1978speckle,wagner1983statistics} that depend on the speckle and tissue characteristics to discriminate tissue types; see \cite{destrempes2010critical,destrempes2013review,yamaguchi2013quantitative} for further references. In either case, quantitative ultrasound can be used to diagnose pathologies with ultrasound images. Another field of application is speckle-tracking that aims at estimating local displacements and tissue deformations; see for instance \cite{ophir1991elastography, hein1993current, lubinski1999speckle}.

On the other hand, the Lorentz force electrical impedance tomography (LFEIT) method \cite{montalibet2002scanning,grasland2013LFEIT}, also known as magneto-acousto electrical tomography \cite{haider2008magneto}, is a medical imaging technique producing electrical conductivity images of tissues \cite{wen1998hall, roth2011role, ammari2014mathematical}. With this technique, an ultrasound wave is transmitted by an acoustic transducer in a biological tissue placed in a magnetic field. The movement of the tissue in the magnetic field due to the ultrasound propagation induces an electric current due to Lorentz force. The measurement of this current with electrodes in contact with the sample allows to reconstruct images of electrical impedance interfaces.

The goal of this work was to assess the presence of an ``acousto-electrical'' speckle in the LFEIT technique, similar to the acoustic speckle in US imaging, as suggested in a previous work \cite{haider2008magneto}. The first part of this work presents the theoretical similarity of measured signals in these two imaging techniques. This similarity is then observed experimentally on an acoustic and electrical interface. Finally, a bovine sample is imaged using both methods to observe the two types of speckle.

\section{Theoretical background}
The goal of this section is to compare the mathematical framework of the LFEIT method with that describing radio-frequency signals in US imaging.

\subsection{Local current density in Lorentz force electrical impedance tomography}
\begin{figure}[!ht]
	\begin{center}
	\includegraphics[width=.8\linewidth]{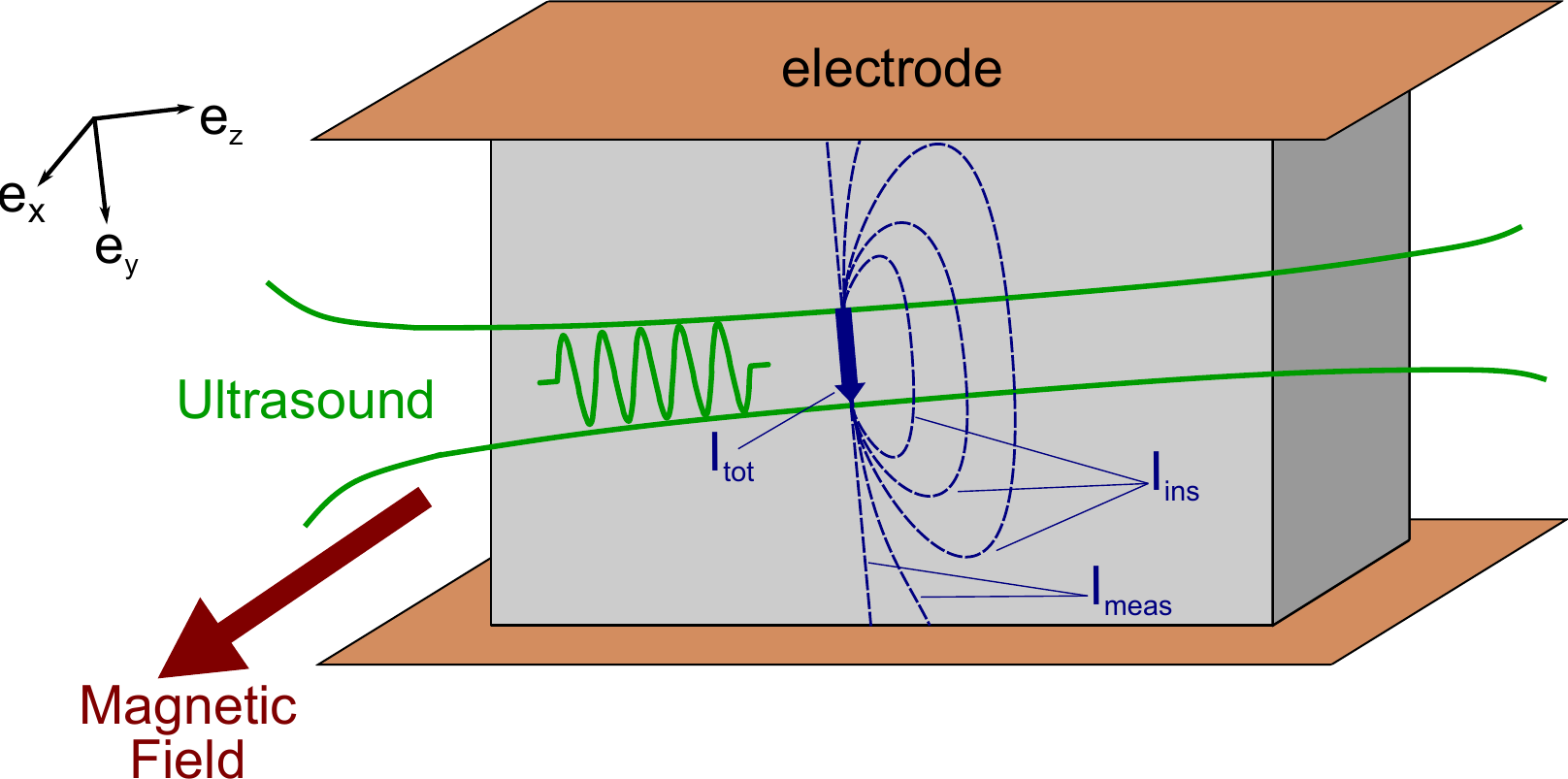}
	\caption{An ultrasound wave is transmitted in a conductive medium placed in a magnetic field. This induces an electric current $I_{tot}$ due to Lorentz force. This current separates in two components, $I_{ins}$ which stays inside the medium, and $I_{meas}$ which is measured by electrodes in contact. The measured signal allows to reconstruct images of electrical impedance interfaces.}
	 \label{Figure1}
	\end{center}
\end{figure}

The LFEIT principle is illustrated in Figure \ref{Figure1}. In this technique, an ultrasound wave is transmitted along a unit vector $\mathbf{e}_z$ in a conductive medium placed in a magnetic field \textcolor{black}{along $\mathbf{e}_x$ of intensity $B_x$}. This induces an electric current due to Lorentz force in the direction $\mathbf{e}_y=\mathbf{e}_z \times \textcolor{black}{\mathbf{e}_x}$. We assume that the conductive medium is moved by the ultrasound wave along the $\mathbf{e}_z$ direction so that all particles within the medium are moved with a mean velocity of amplitude $v_{\textcolor{black}{z}}$ parallel to $\mathbf{e}_z$ \cite{mari2009bulk}. Since the medium is placed in a magnetic field $B_x$, a particle $k$ of charge $q_k$ is deviated by a Lorentz force $\mathbf{F}_k=q_k v_{\textcolor{black}{z}} B_x \mathbf{e}_y$. Using Newton's second law, the velocity $\mathbf{u}_k$ of the charged particle can be calculated as:
\begin{equation}
\mathbf{u}_k = v_{\textcolor{black}{z}} \mathbf{e}_z + \mu_k v_{\textcolor{black}{z}} B_x \mathbf{e}_y,
\label{eq1}
\end{equation}
where $\mu_k$ is the mobility of the particle $k$.

The density of current $\mathbf{j}$, defined as $\sum_k q_k \mathbf{u}_k$ \textcolor{black}{per unit volume}, is then equal to the sum of $\sum_k q_k v_{\textcolor{black}{z}} \mathbf{e}_z$ and $\sum_k q_k \mu_k v_{\textcolor{black}{z}} B_x \mathbf{e}_y$ \textcolor{black}{per unit volume}. If the medium is assumed electrically neutral, i.e. $\sum_k q_k=0$, the first term is equal to zero. By introducing the electrical conductivity \textcolor{black}{$\sigma$}, defined as $\sum_k q_k \mu_k$ \textcolor{black}{per unit volume}, the second term is equal to $\sigma v_{\textcolor{black}{z}} B_x \mathbf{e}_y$. Equation (\ref{eq1}) can consequently be written as:
\begin{equation}
\mathbf{j}=\sigma v_{\textcolor{black}{z}} B_x \mathbf{e}_y.
\label{eq2}
\end{equation}
The local density of current $\mathbf{j}$ is thus proportional to the electrical conductivity of the sample $\sigma$, to the component $B_x$ of the applied magnetic field, and to the local sample speed $v$ \cite{montalibet2002scanning}.

\subsection{Measured signal in Lorentz force electrical impedance tomography}
Equation (\ref{eq2}) is however local. We consider, as a first approximation, the ultrasound beam as a plane wave along $\mathbf{e}_z$ \textcolor{black}{inside a disk of diameter $W$}, with the velocity $v_{\textcolor{black}{z}} = 0$ outside the beam. The total current $I_{tot}$ induced by the Lorentz force is then equal to the \textcolor{black}{average} of the current density $j$ \textcolor{black}{computed over all surfaces $S$} inside the ultrasound beam that are perpendicular to $e_y$ \cite{montalibet2002these}:
\begin{equation}
		I_{tot} (t) = \frac{1}{\textcolor{black}{W}} \textcolor{black}{\int}{\int{\int{\mathbf{j}.\mathbf{dS} \textcolor{black}{~dy}}}} =\frac{1}{\textcolor{black}{W}}\textcolor{black}{ \int}{\int{\int{\sigma v_{\textcolor{black}{z}} B_x ~dz ~dx \textcolor{black}{~dy}}}}.
\label{eq3}
\end{equation}
As electrodes are placed outside the beam, we assume they measure only a fraction $\alpha$ of the total current $I_{tot}$ induced by the Lorentz force. The current $I_{tot}$ then induces a current outside the beam which separates in two components: a current $I_{ins}$ that stays inside the medium and operates through a resistance $R_{ins}$, and a current $I_{meas}$ that is measured by electrodes and operates through a resistance $R_{meas}$. Due to Ohm's law, which states that $I_{ins} R_{ins} = I_{meas} R_{meas}$, the coefficient $\alpha$, defined as $I_{meas}/I_{tot}$, is also equal to $R_{ins}/(R_{meas} + R_{ins})$. This coefficient, rather difficult to estimate, depends consequently on a few parameters: the size and location of electrodes, and the circuit and medium electrical impedance. We can nevertheless conclude that the lower the resistance $R_{meas}$ compared to $R_{ins}$, the higher $\alpha$ is, and the higher is the measured electrical current.

For a linear and inviscid medium, the medium velocity $v_{\textcolor{black}{z}}$ due to ultrasound wave propagation is related to the ultrasound pressure $p$ and the medium density $\rho$ (assumed here constant over time but not necessarily over \textcolor{black}{space}) using the identity $\frac{\partial v_{\textcolor{black}{z}} (t,z)}{\partial t} = - \frac{1}{\rho}\frac{\partial p(t,z)}{\partial z}$ \textcolor{black}{(having assumed a plane wave)}. From these considerations on the coefficient $\alpha$ and the medium velocity $v_{\textcolor{black}{z}}$, the measured current $I_{meas}$ is equal to:
\begin{equation}
	I_{meas}(t) = \frac{\alpha}{\textcolor{black}{W}} \textcolor{black}{\int}{\textcolor{black}{\int}{\int_{z_1}^{z_2}{\sigma B_x \Bigl (\int_{-\infty}^{t}{-\frac{1}{\rho}\frac{\partial p(\tau,z)}{\partial z}~d\tau} \Bigr ) ~dz\textcolor{black}{~dx~dy}}}},
\label{eq4}
\end{equation}
where $z_1$ and $z_2$ are the boundaries of the studied medium along $\mathbf{e}_z$.

Considering a progressive acoustic wave that propagates only along the direction $\mathbf{e}_z$ and is not attenuated in the measurement volume, we may assume that $p(\tau,z) $ is of the form $P(\tau-z/c)$ where $c$ is the speed of sound in the medium. Equation \ref{eq4} can consequently be written as:
\begin{equation}
	I_{meas}(t) = \frac{\alpha}{\textcolor{black}{W}} \textcolor{black}{\int}{\textcolor{black}{\int}{\int_{z_1}^{z_2}{\sigma B_x \frac{1}{\rho c} \Bigl ( P(t-z/c) - P(-\infty) \Bigr ) ~dz\textcolor{black}{~dx~dy}}}},
\label{eq5}
\end{equation}
where $P(-\infty) = 0$ since the transmitted pulse is of finite length.

Assuming $B_x$ and \textcolor{black}{$c$ constant over space} ($c$ varies typically from -5 to +5\% between soft biological tissues \cite{hill2004physical}), and replacing the integration over $z$ by an integration over $\tau = z/c$, equation \ref{eq5} becomes:
\begin{equation}
	I_{meas}(t) = \frac{\alpha B_x}{\textcolor{black}{W}} \int_{-\infty}^{\infty}{H_{\textcolor{black}{z}}(\tau) P(t-\tau) ~d\tau},
\label{eq6}
\end{equation}
where \textcolor{black}{$H_z(\tau)=\int{\int{H(x^\prime,y^\prime,z^\prime=c\tau)dx^\prime dy^\prime}}$, $H$ is equal to $\frac{\sigma}{\rho}$ within the studied medium and $0$ elsewhere.} Since the medium density vary typically by a few percent between different soft tissues \cite{cobbold2007foundations}, while the electrical conductivity of soft tissues can vary up to a few tens \cite{gabriel1996dielectric2}, $H$ can be seen mostly as a variable describing the electrical conductivity of the tissue. One can show that if the DC component of the transmitted pressure is null, then $I_{meas}$ is null whenever the electrical conductivity is constant \cite{wen1998hall}. Thus, equation \ref{eq6} shows that the measured electric current is proportional to the convolution product of a function $H_{\textcolor{black}{z}}$ of the electrical conductivity with the axial transmitted pressure wave $P$.

\subsection{Measured signal in B-mode ultrasound imaging}
Ultrasound imaging is based on the measurement of reflections of the transmitted acoustic wave (i.e., backscattering, diffraction or specular reflection). We assume that acoustic inhomogeneities are scattering only a small part of the transmitted pressure, so that scattered waves have a negligible amplitude compared to the main acoustic beam, and that the diffraction and attenuation are small in the region of interest. The voltage $RF(t)$ measured at time $t$ on the transducer, known as the radiofrequency signal, is then equal to:
\begin{equation}
   RF(t)=D\int_{-\infty}^{\infty}{T_{z}(\tau) P(t-\tau)~d\tau},
	\label{eq8}
\end{equation}
where $D$ is a constant of the transducer related to the acousto-electric transfer function, $T_{z}(\tau)=\int{\int{ T(x^\prime,y^\prime,z^\prime=c\tau\textcolor{black}{/2})~dx^\prime~dy^\prime}}$ with $T$ the continuous spatial distribution of point scatterers, $P(t)$ is the axial transmitted pressure wave (the pulse shape), and $\tau=\textcolor{black}{2}z/c$ \textcolor{black}{(the factor $2$ takes into account round-trip wave propagation in ultrasound imaging)}. In the latter expression of $T_z(\tau)$, the double integral is performed over a disk of diameter $W$ (the ultrasound beam is considered as a plane wave along $\mathbf{e}_z$ inside a limited width $W$ as in the previous section) \cite{bamber1980ultrasonic}. The envelope of $RF(t)$ represents the A-mode ultrasound signal. \textcolor{black}{Note that under a more realistic incident pressure wave model, one can consider pulse beam profiles in the definition of $H_z$ (at emission) and $T_z$ (at emission and at reception), which can be deduced by multiplying the pulse shape with the incident pulse beam profile}. Also, attenuation can be taken into account by convolution of Eq. \ref{eq8} (and similarly, of Eq. \ref{eq6}) with an attenuation function.

Hence, the received signal in US imaging is proportional to the convolution product of the axial spatial distribution of point scatterers $T_z$ with the transmitted pulse shape $P$. If scatterers are small compared to the ultrasound wavelength, reflected waves can interfere. When the envelope of the signal is calculated and a B-mode image is formed line by line, this phenomenon appears on the image as a granular texture called acoustic speckle \cite{abbott1979acoustic}.

Apart from a proportionality coefficient, equations (\ref{eq6}) and (\ref{eq8}) have a similar form, where $T$ and $H$ play an analogous role. A speckle phenomenon is thus expected to be observed in the LFEIT technique. Its nature would be electrical, because mainly related to electrical conductivity inhomogeneities, and also acoustic, because spatial characteristics are related to the acoustic wavelength.

\section{Materials and methods}
Two experiments were performed in this study. The first experiment aimed at comparing a LFEIT signal and a US signal on a simple acoustic and electrical conductivity interface using the same acoustic transducer. This approach was used to perform a test with a large change in both electrical conductivity and acoustic impedance. Thus, tissue functions $T_{\textcolor{black}{z}}$ and $H_{\textcolor{black}{z}}$ could be considered as square pulse functions with strong gradients at the interface locations. According to the \textcolor{black}{above hypotheses}, the two \textcolor{black}{scan line} signals are expected to be identical apart from a proportionality coefficient \textcolor{black}{in this special case}.

The goal of the second experiment was to observe a complex biological tissue with the two imaging techniques and using the same acoustic transducer. A speckle pattern of similar spatial characteristics is expected to be observed on both types of images, since they are related to the acoustic wavelength and beam width, but different, because the nature of the imaged parameter is acoustic in one case and electrical in the other case.

\subsection{Measured signal on an acousto-electrical interface}
\begin{figure}[!ht]
	\begin{center}
 \includegraphics[width=1\linewidth]{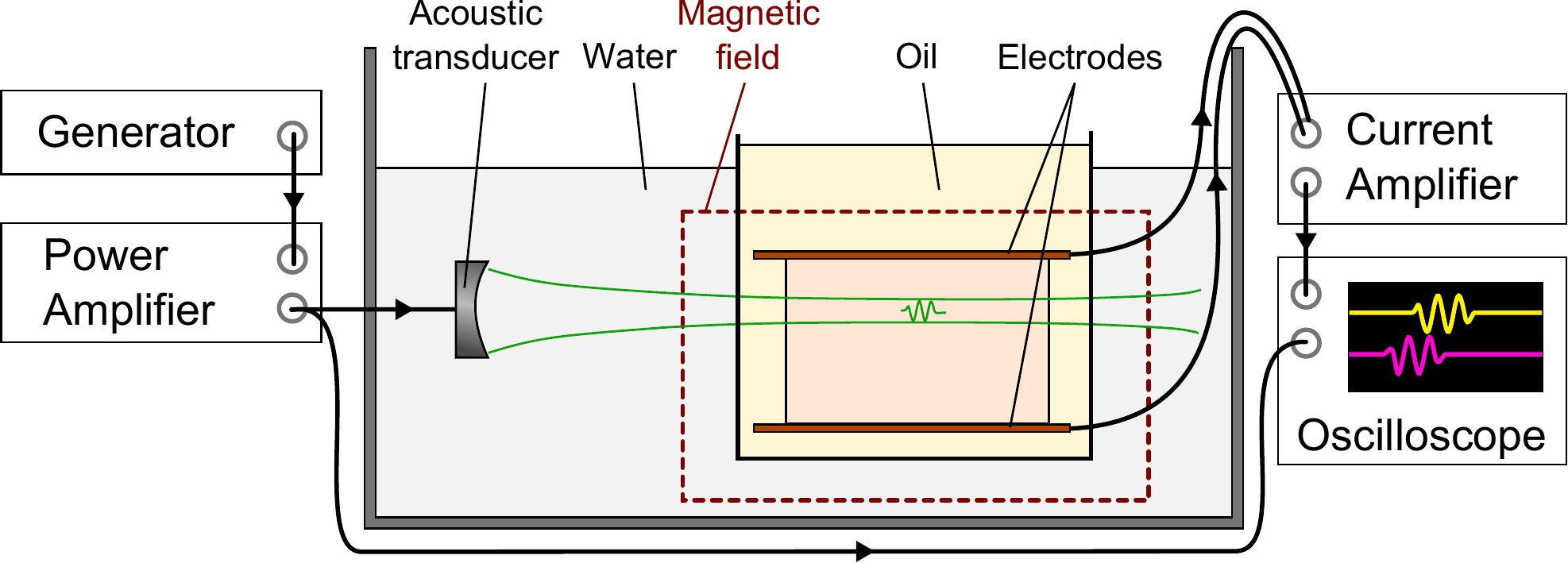}
	\caption{A transducer is transmitting ultrasound pulses toward a sample placed in an oil tank placed in a magnetic field. The induced electric current is received by two electrodes in contact respectively with two sides of the gelatin.}
	 \label{Figure2}
	\end{center}
\end{figure}

The experiment setup is illustrated in Figure \ref{Figure2}. A generator (HP33120A, Agilent, Santa Clara, CA, USA) was used to create 0.5 MHz, 3 cycles sinusoid bursts at a pulse repetition frequency of 100 Hz. This excitation was amplified by a 200 W linear power amplifier (200W LA200H, Kalmus Engineering, Rock Hill, SC, USA) and sent to a 0.5 MHz, 50 mm in diameter transducer focused at 210 mm and placed in a degassed water tank. The peak-to-peak pressure at the focal point was equal to 3 MPa. A 4x15x20 cm$^3$ mineral oil tank was located from 15 to 35 cm away from the transducer in the ultrasound beam axis. This oil tank was used to decrease the loss of current from the sample to the surrounding medium and to increase consequently the signal-to-noise ratio, but was not mandatory. It was inserted in a 300$\pm$50 mT magnetic field created by a U-shaped permanent magnet, composed of two poles made of two 3x5x5 cm$^3$ NdFeB magnets (BLS Magnet, Villers la Montagne, France) separated by a distance of 4.5 cm.

The tested medium was a 5x10x10 cm$^3$ 10\% gelatin filled with 5\% salt sample placed inside the oil tank from 30 to 40 cm away from the transducer, and presented consequently a strong acoustic and electrical interface. A pair of 3x0.1x10 cm$^3$ copper electrodes was placed in contact with the sample, above and under it, respectively. The electrodes were linked through an electrical wire to a 1 MV/A current amplifier (HCA-2M-1M, Laser Components, Olching, Germany). A voltage amplifier could also be used, but the current amplifier presents a smaller input impedance (a few Ohms vs 50 Ohms) and increases consequently the amount of current measured by the electrode, as depicted by the factor $\alpha$ in equation \ref{eq4}. The signals were then measured by an oscilloscope with 50 $\Omega$ input impedance (WaveSurfer 422, LeCroy, Chestnut Ridge, NY, USA) and averaged over 1000 acquisitions. US signals were simultaneously recorded using the same oscilloscope with a 1/100 voltage probe.

To quantify similarities between the two signals, we computed the correlation coefficient between them. Such coefficient is equal to 1 when the two signals are directly proportional and 0 when they are uncorrelated.

\subsection{Observation of speckle in a biological tissue}
\begin{figure}[!ht]
	\begin{center}
	\includegraphics[width=.5\linewidth]{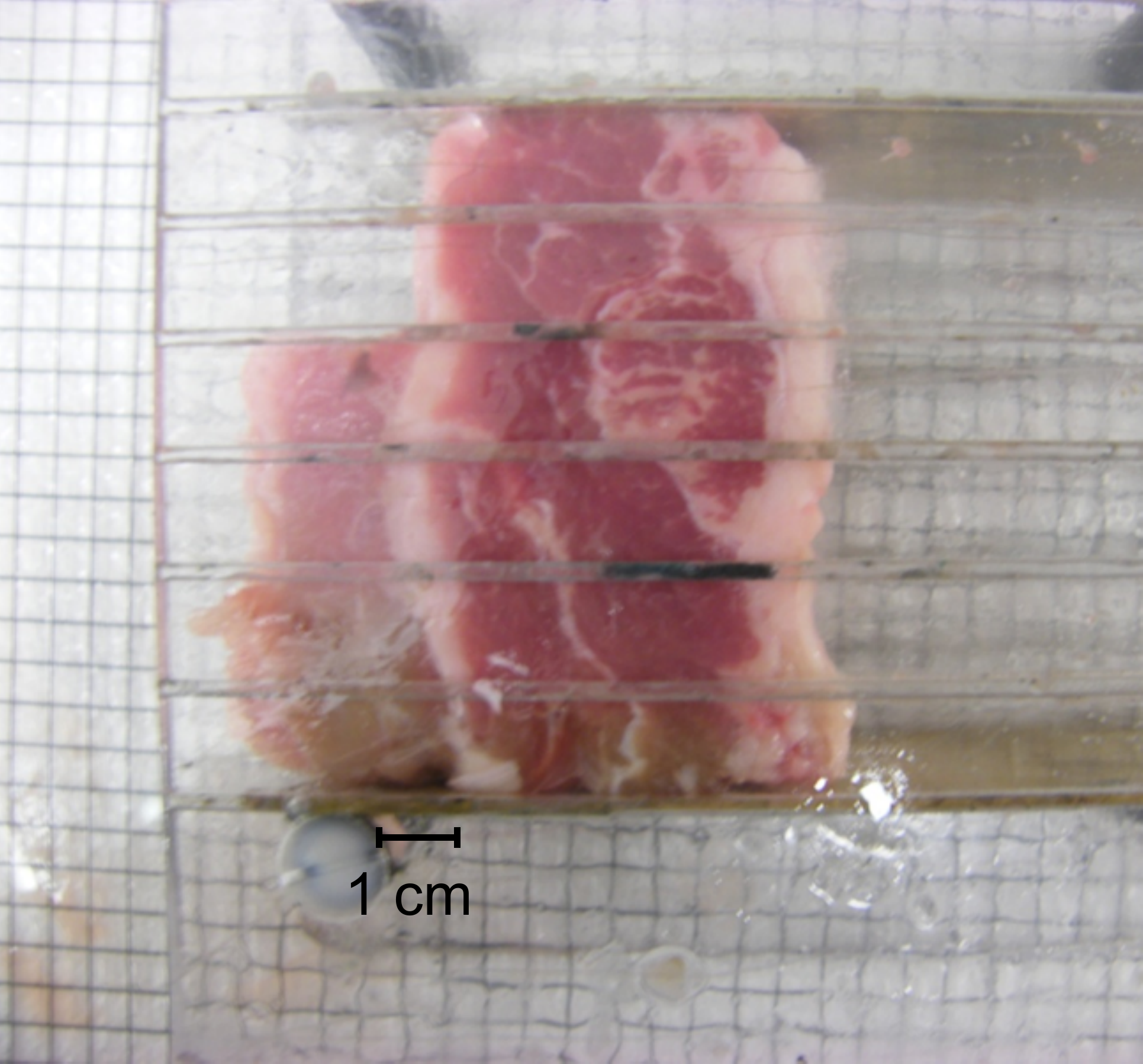}
	\caption{Picture of the 2x6x6 cm$^3$ imaged bovine rib muscle sample, with many fat inclusions.}
	 \label{Figure3}
	\end{center}
\end{figure}

The same apparatus as in the first experiment was used, but the gelatin sample was replaced by a 2x6x6 cm$^3$ piece of bovine rib muscle purchased at a grocery store. It presented many fat inclusions, as pictured in Figure \ref{Figure3}. B-mode images were produced line by line by moving the transducer along the $e_y$ direction by 96 steps of 0.5 mm. Acoustic and electrical signals were post-processed using the Matlab software (The MathWorks, Natick, MA, USA) by computing the magnitude of the Hilbert transform of the signal \cite{roden1991analog}, and were displayed with grayscale and jet color images, respectively.


\section{Results}
\begin{figure}[!ht]
	\begin{center}
	\includegraphics[width=1\linewidth]{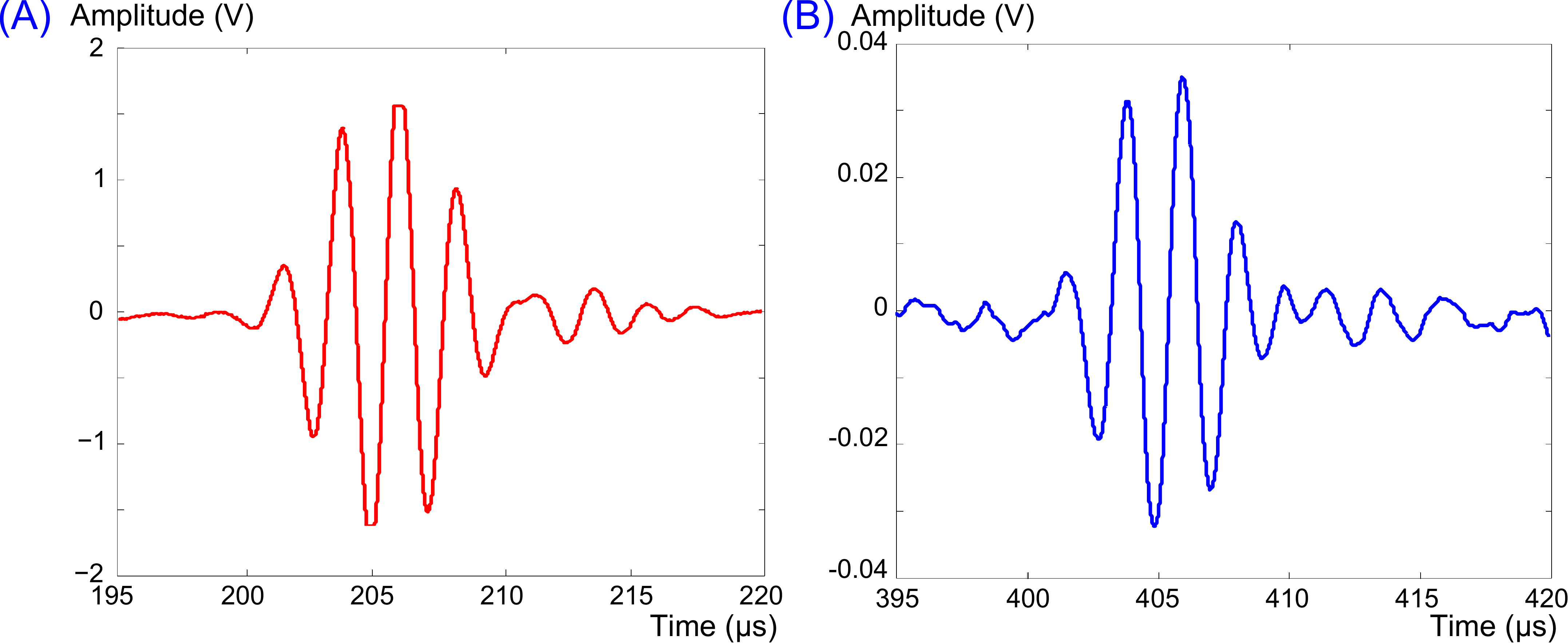}
	\caption{(A) Electrical signal acquired by electrodes from the first phantom interface, 195 to 220 $\mathrm{\mu}$s after acoustic wave transmission. The signal is made of three to four cycles at 500 kHz. (B) Electrical signal acquired by the acoustic transducer from the first phantom interface, 395 to 420 $\mathrm{\mu}$s after acoustic wave transmission. The signal is also made of three to four cycles at 500 kHz.}
	 \label{Figure4}
	\end{center}
\end{figure}

Figure \ref{Figure4}-(A) presents the electrical signal measured by electrodes from the first phantom interface, 195 to 220 microseconds after acoustic transmission, corresponding to a distance of 30 cm. Figure \ref{Figure4}-(B) depicts the signal acquired by the US transducer from the phantom interface, 395 to 420 microseconds after acoustic transmission, corresponding to a distance of 30 cm (back and forth). Both signals consist in three to four cycles at a central frequency of 500 kHz. The correlation coefficient between both signals is 0.9297, indicating a high similarity. This similarity was observed at the second phantom interface, with a correlation coefficient between both signals equal to 0.9179. LFEIT and US images of the phantom can be seen in \cite{grasland2013LFEIT}.

\begin{figure}[!ht]
	\begin{center}
	 \includegraphics[width=1\linewidth]{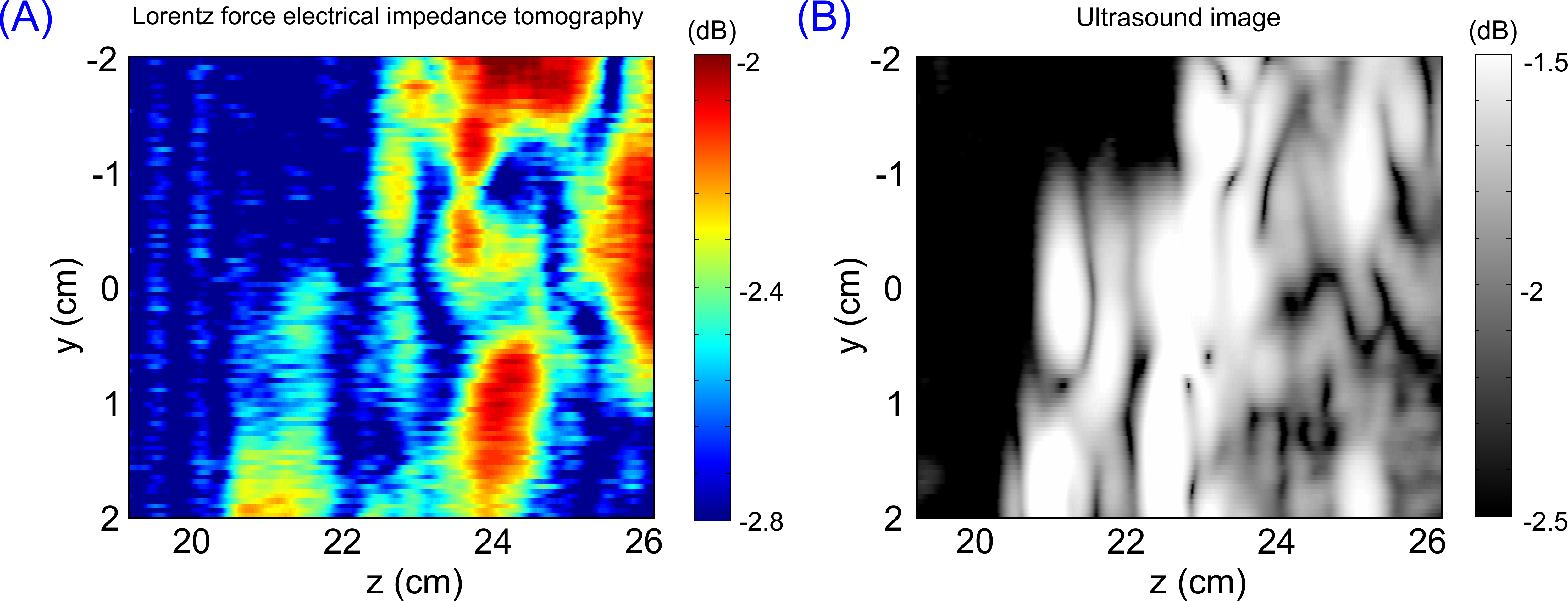}
 	\caption{(A) Lorentz force electrical impedance tomography (LFEIT) image of the bovine muscle sample. (B) Ultrasound (US) image of the same bovine rib muscle sample. A speckle pattern can be seen inside the medium.}
 	\label{Figure5}
	\end{center}
\end{figure}

Figures \ref{Figure5}-(A) and -(B) present the Lorentz force electrical impedance tomography image and the ultrasound image, respectively, of the bovine muscle sample. The amplitude varied from -2 to -2.8 dB with the first technique and from -1.5 to -2.5 dB with the other (0 dB being a measured amplitude of 1 V). Main interfaces of the medium can be retrieved on both images, as previously shown \cite{grasland2013LFEIT}, \textcolor{black}{even if signals at the boundaries  and the  interior of the bovine sample were of similar amplitude}. A speckle pattern was present in both images. \textcolor{black}{The typical spots were of size of same order of magnitude; i.e., 5-8 mm in the Z-direction and 10-18 mm in the Y-direction, but their spatial distributions were different, as expected due to the difference between the electrical and acoustic inhomogeneities. For each image, the signal-to-noise ratio was estimated as the base 10 logarithm of the ratio of the mean square amplitude of the RF signals  in the 22.5-26 cm zone with the mean square amplitude in the 19-20 cm noisy zone. The signal-to-noise ratio was lower by 0.9 dB in the LFEIT image compared to the US image.}

\section{Discussion}
The gelatin phantom used in the first experiment presented an interface of acoustic and electrical impedances. According to the correlation coefficient, the measured signals were very similar, which is a first indication of the validity of the approach presented in the theoretical section: the reflected wave is proportional to the convolution product of the acoustic impedance distribution with the transmitted ultrasound pulse shape, while the induced electric current is proportional to the convolution product of the electrical impedance distribution with the transmitted ultrasound pulse shape. 

The second experiment showed two images of granular pattern. This pattern does not represent macroscopic variations of acoustic or electrical impedances, and we interpreted it as speckle. The granular pattern appeared visually with similar characteristics of size and shape in both images. Note that the size of the speckle spots along the ultrasound beam was different than in the orthogonal direction, because the first is mainly related to the acoustic wavelength and the second to the ultrasound beam width \cite{obrien1993single}. The spots were quite larger than those usually seen in clinical ultrasound images, due to the characteristics of the instrument (three cycles at 500 kHz, 1.5 cm beam width).

Although the bright spots were of similar size, their precise locations in the two images were different. This shows that the observed speckle reveals information of different nature in the two modalities: acoustic or electrical inhomogeneities. In this case, the bovine meat sample presented not only large layers of fat but also small inclusions of adipose tissues whose electrical conductivity differs from muscle (differences can be ten time higher at 500 kHz \cite{gabriel1996dielectric2}), whereas they have a close acoustic impedance (differences smaller than 10\% \cite{cobbold2007foundations}).

We introduce henceforth the term ``acousto-electrical speckle''. It is justified by the fact that its spatial characteristics are related to the acoustic parameters, especially the ultrasound wavelength, while its nature is related to the electrical impedance variation distribution $H$. The existence of this speckle could allow using speckle-based ultrasound techniques in the Lorentz force electrical impedance tomography technique, for example compound imaging, speckle-tracking algorithm or quantitative ultrasound for tissue characterization purposes \cite{odonnell1994internal,jespersen1998multi,mamou2013quantitative}. These techniques have however not be applied in this study because of the low spatial resolution of the images due to the low frequency transducer used.

This study shows that the electrical impedance inhomogeneities can be studied using LFEIT at a scale controlled by the acoustic wavelength instead of the electromagnetic wavelength, which is 5 orders of magnitude larger at a same frequency, and would be prohibitive in the context of biological tissues imaging.

These inhomogeneities should also be observed in a ``reverse'' mode (terminology from Wen et al. \cite{wen1998hall}) called Magneto-Acoustic Tomography with Magnetic Induction, where an electrical current and a magnetic field are combined to produce ultrasound waves \cite{roth1994theoretical, xu2005magneto, ammari2009mathematical}. However, presence of speckle in this last technique has not yet been demonstrated and which of the two methods will be most useful in biomedical imaging is not clear \cite{roth2011role}.

\section{Conclusion}
In this study, the similarity between two imaging modalities, the Lorentz force electrical impedance tomography and ultrasound imaging, was assessed theoretically. This similarity was then observed experimentally on a basic acoustic and electrical interface with both methods. Then, the two techniques were used to image a biological tissue presenting many acoustic and electrical impedance inhomogeneities. The speckle pattern formed in both images exhibited similar spatial characteristics. This suggests the existence of an ``acousto-electrical speckle'' with spatial characteristics driven by acoustic parameters but due to electrical impedance variation distribution. This allows considering the use of ultrasound speckle-based image processing techniques on Lorentz force electrical impedance tomography data and to study electrical inhomogeneity structures at ultrasound wavelength scale.

\section{Acknowledgments}
Part of the project was financed by a Discovery grant of the Natural Sciences and Engineering Research Council of Canada (grant \# 138570-11). The first author was recipient of a post-doctoral fellowship award by the FRM SPE20140129460 grant. The authors declare no conflict of interest in the work presented here.

\section{Bibliography}
\bibliographystyle{apalike}
\bibliography{biblio}

\begin{thebibliography}{}

\bibitem[Abbott and Thurstone, 1979]{abbott1979acoustic}
Abbott, J.~G. and Thurstone, F. (1979).
\newblock {Acoustic speckle: Theory and experimental analysis}.
\newblock {\em Ultrasonic Imaging}, 1(4):303--324.

\bibitem[Ammari et~al., 2009]{ammari2009mathematical}
Ammari, H., Capdeboscq, Y., Kang, H., Kozhemyak, A., et~al. (2009).
\newblock {Mathematical models and reconstruction methods in magneto-acoustic
  imaging}.
\newblock {\em European Journal of Applied Mathematics}, 20:303--317.

\bibitem[Ammari et~al., 2014]{ammari2014mathematical}
Ammari, H., Grasland-Mongrain, P., Millien, P., Seppecher, L., and Seo, J.-K.
  (2014).
\newblock {A mathematical and numerical framework for ultrasonically-induced
  Lorentz force electrical impedance tomography}.
\newblock {\em Journal de Mathématiques Pures et Appliquées}, submitted.

\bibitem[Bamber and Dickinson, 1980]{bamber1980ultrasonic}
Bamber, J. and Dickinson, R. (1980).
\newblock {Ultrasonic B-scanning: a computer simulation}.
\newblock {\em Physics in Ultrasound in Medicine and Biology}, 25(3):463--479.

\bibitem[Burckhardt, 1978]{burckhardt1978speckle}
Burckhardt, C.~B. (1978).
\newblock Speckle in ultrasound b-mode scans.
\newblock {\em IEEE Transactions on Sonics and Ultrasonics}, 25(1):1--6.

\bibitem[Cobbold, 2007]{cobbold2007foundations}
Cobbold, R.~S. (2007).
\newblock {\em {Foundations of Biomedical Ultrasound}}.
\newblock Oxford University Press, USA.

\bibitem[Destrempes and Cloutier, 2010]{destrempes2010critical}
Destrempes, F. and Cloutier, G. (2010).
\newblock {A critical review and uniformized representation of statistical
  distributions modeling the ultrasound echo envelope}.
\newblock {\em Ultrasound in Medicine and Biology}, 36(7):1037--1051.

\bibitem[Destrempes and Cloutier, 2013]{destrempes2013review}
Destrempes, F. and Cloutier, G. (2013).
\newblock {Review of envelope statistics models for quantitative ultrasound
  imaging and tissue characterization}.
\newblock In Mamou, J. and Oelze, M.~L., editors, {\em {Quantitative Ultrasound
  in Soft Tissues}}, pages 219--274. Springer Dordrecht Heidelberg New York
  London.

\bibitem[Gabriel et~al., 1996]{gabriel1996dielectric2}
Gabriel, S., Lau, R., and Gabriel, C. (1996).
\newblock {The dielectric properties of biological tissues: II. Measurements in
  the frequency range 10 Hz to 20 GHz}.
\newblock {\em Physics in Medicine and Biology}, 41(11):2251--2269.

\bibitem[Goshal et~al., 2013]{goshal2013state}
Goshal, G., Mamou, J., and Oelze, M.~L. (2013).
\newblock {State of the art methods for estimating backscatter coefficients}.
\newblock In Mamou, J. and Oelze, M.~L., editors, {\em {Quantitative Ultrasound
  in Soft Tissues}}, pages 3--19. Springer Dordrecht Heidelberg New York
  London.

\bibitem[Grasland-Mongrain et~al., 2013]{grasland2013LFEIT}
Grasland-Mongrain, P., Mari, J.-M., Chapelon, J.-Y., and Lafon, C. (2013).
\newblock {Lorentz force electrical impedance tomography}.
\newblock {\em Innovation and Research in Biomedical Engineering},
  34(4-5):357--360.

\bibitem[Haider et~al., 2008]{haider2008magneto}
Haider, S., Hrbek, A., and Xu, Y. (2008).
\newblock {Magneto-acousto-electrical tomography: a potential method for
  imaging current density and electrical impedance}.
\newblock {\em Physiological Measurement}, 29(6):S41--S50.

\bibitem[Hein and O'Brien, 1993]{hein1993current}
Hein, I. and O'Brien, W.~D. (1993).
\newblock {Current time-domain methods for assessing tissue motion by analysis
  from reflected ultrasound echoes-a review}.
\newblock {\em IEEE Transactions on Ultrasonics, Ferroelectrics and Frequency
  Control}, 40(2):84--102.

\bibitem[Hill et~al., 2004]{hill2004physical}
Hill, C.~R., Bamber, J.~C., and Haar, G. (2004).
\newblock {\em {Physical principles of medical ultrasonics}}, volume~2.
\newblock Wiley Online Library.

\bibitem[Insana et~al., 1990]{insana1990describing}
Insana, M.~F., Wagner, R.~F., Brown, D.~G., and Hall, T.~J. (1990).
\newblock {Describing small-scale structure in random media using pulse-echo
  ultrasound}.
\newblock {\em Journal of Acoustical Society of America}, 87(1):179--192.

\bibitem[Jespersen et~al., 1998]{jespersen1998multi}
Jespersen, S.~K., Wilhjelm, J.~E., and Sillesen, H. (1998).
\newblock {Multi-angle compound imaging}.
\newblock {\em Ultrasonic Imaging}, 20(2):81--102.

\bibitem[Lizzi et~al., 1983]{lizzi1983theoretical}
Lizzi, F.~L., Greenabaum, M., Feleppa, E.~J., Elbaum, M., and Coleman, D.~J.
  (1983).
\newblock {Theoretical framework for spectrum analysis in ultrasonic tissue
  characterization}.
\newblock {\em Journal of Acoustical Society America}, 73(4):1366--1373.

\bibitem[Lizzi et~al., 1987]{lizzi1987relationship}
Lizzi, F.~L., Ostromogilsky, M., Feleppa, E.~J., Rorke, M.~C., and Yaremko,
  M.~M. (1987).
\newblock {Relationship of ultrasonic spectral parameters to features of tissue
  microstructure}.
\newblock {\em IEEE Transactions on Ultrasonics, Ferroelectrics and Frequency
  Control}, 34(3):319--329.

\bibitem[Lubinski et~al., 1999]{lubinski1999speckle}
Lubinski, M.~A., Emelianov, S.~Y., and O'Donnell, M. (1999).
\newblock {Speckle tracking methods for ultrasonic elasticity imaging using
  short-time correlation}.
\newblock {\em IEEE Transactions on Ultrasonics, Ferroelectrics and Frequency
  Control}, 46(1):82--96.

\bibitem[Mamou and Oelze, 2013]{mamou2013quantitative}
Mamou, J. and Oelze, M.~L. (2013).
\newblock {\em {Quantitative Ultrasound in Soft Tissues}}.
\newblock Springer Dordrecht Heidelberg New York London.

\bibitem[Mari et~al., 2009]{mari2009bulk}
Mari, J.~M., Blu, T., Matar, O.~B., Unser, M., and Cachard, C. (2009).
\newblock {A bulk modulus dependent linear model for acoustical imaging}.
\newblock {\em Journal of Acoustical Society of America}, 125:2413--2419.

\bibitem[Montalibet, 2002]{montalibet2002these}
Montalibet, A. (2002).
\newblock {\em {Etude du couplage acousto-magn{\'e}tique: d{\'e}tection des
  gradients de conductivit{\'e} {\'e}lectrique en vue de la caract{\'e}risation
  tissulaire}}.
\newblock PhD thesis, Institut Nationale des Sciences Appliqu{\'e}es de Lyon.

\bibitem[Montalibet et~al., 2001]{montalibet2002scanning}
Montalibet, A., Jossinet, J., and Matias, A. (2001).
\newblock {Scanning electric conductivity gradients with ultrasonically-induced
  Lorentz force}.
\newblock {\em Ultrasonic Imaging}, 23(2):117--132.

\bibitem[Noble and Boukerroui, 2006]{noble2006ultrasound}
Noble, J.~A. and Boukerroui, D. (2006).
\newblock {Ultrasound image segmentation: a survey}.
\newblock {\em IEEE Transactions on Medical Imaging}, 25(8):987--1010.

\bibitem[O'Brien, 1993]{obrien1993single}
O'Brien, W. (1993).
\newblock {Single-element transducers.}
\newblock {\em Radiographics}, 13(4):947--957.

\bibitem[O'Donnell et~al., 1994]{odonnell1994internal}
O'Donnell, M., Skovoroda, A.~R., Shapo, B.~M., and Emelianov, S.~Y. (1994).
\newblock {Internal displacement and strain imaging using ultrasonic speckle
  tracking}.
\newblock {\em IEEE Transactions on Ultrasonics Ferroelectrics and Frequency
  Control}, 41(3):314--325.

\bibitem[Ophir et~al., 1991]{ophir1991elastography}
Ophir, J., Cespedes, I., Ponnekanti, H., Yazdi, Y., and Li, X. (1991).
\newblock {Elastography: a quantitative method for imaging the elasticity of
  biological tissues}.
\newblock {\em Ultrasonic Imaging}, 13(2):111--134.

\bibitem[Roden, 1991]{roden1991analog}
Roden, M.~S. (1991).
\newblock {\em {Analog and Digital Communication Systems}}, volume~1.
\newblock Englewood Cliffs, NJ, Prentice Hall.

\bibitem[Roth, 2011]{roth2011role}
Roth, B.~J. (2011).
\newblock {The role of magnetic forces in biology and medicine}.
\newblock {\em Experimental Biology and Medicine}, 236(2):132--137.

\bibitem[Roth et~al., 1994]{roth1994theoretical}
Roth, B.~J., Basser, P.~J., and Wikswo, J.~P. (1994).
\newblock {A theoretical model for magneto-acoustic imaging of bioelectric
  currents}.
\newblock {\em IEEE Transactions on Biomedical Engineering}, 41(8):723--728.

\bibitem[Wagner et~al., 1983]{wagner1983statistics}
Wagner, R.~F., Smith, S.~W., Sandrik, J.~M., and Lopez, H. (1983).
\newblock {Statistics of speckle in ultrasound B-scans.}
\newblock {\em IEEE Transactions on Sonics and Ultrasonics}, 30(3):156--163.

\bibitem[Wen et~al., 1998]{wen1998hall}
Wen, H., Shah, J., and Balaban, R.~S. (1998).
\newblock {Hall effect imaging}.
\newblock {\em IEEE Transactions on Biomedical Engineering}, 45(1):119--124.

\bibitem[Xu and He, 2005]{xu2005magneto}
Xu, Y. and He, B. (2005).
\newblock {Magnetoacoustic tomography with magnetic induction (MAT-MI)}.
\newblock {\em Physics in Medicine and Biology}, 50(21):5175--5192.

\bibitem[Yamaguchi, 2013]{yamaguchi2013quantitative}
Yamaguchi, T. (2013).
\newblock {The quantitative ultrasound diagnosis of liver fibrosis using
  statistical analysis of the echo envelope}.
\newblock In Mamou, J. and Oelze, M.~L., editors, {\em {Quantitative Ultrasound
  in Soft Tissues}}, pages 275--288. Springer Dordrecht Heidelberg New York
  London.

\end{thebibliography}

\end{document}